\documentclass[12pt, onecolumn]{IEEEtran}
\renewcommand{\baselinestretch}{1.5}

\usepackage{graphicx}

\newtheorem{theorem}{\bf Theorem}

\newtheorem{lemma}{\bf Lemma}

\newtheorem{assumption}{\bf Assumption}

\begin{document}
%
\title{On Low Complexity Maximum Likelihood Decoding of Convolutional Codes}
%
%
\author{Jie~Luo, ~\IEEEmembership{Member,~IEEE}
\thanks{The author is with the Electrical and Computer Engineering Department, Colorado State University, Fort Collins, CO 80523. E-mail: rockey@engr.colostate.edu. }
\thanks{This work was supported by National Science Foundation grant CCF-0728826.}
}

\maketitle

\begin{abstract}
This paper considers the average complexity of maximum likelihood (ML) decoding of convolutional codes. ML decoding can be modeled as finding the most probable path taken through a Markov graph. Integrated with the Viterbi algorithm (VA), complexity reduction methods such as the sphere decoder often use the sum log likelihood (SLL) of a Markov path as a bound to disprove the optimality of other Markov path sets and to consequently avoid exhaustive path search. In this paper, it is shown that SLL-based optimality tests are inefficient if one fixes the coding memory and takes the codeword length to infinity. Alternatively, optimality of a source symbol at a given time index can be testified using bounds derived from log likelihoods of the neighboring symbols. It is demonstrated that such neighboring log likelihood (NLL)-based optimality tests, whose efficiency does not depend on the codeword length, can bring significant complexity reduction to ML decoding of convolutional codes. The results are generalized to ML sequence detection in a class of discrete-time hidden Markov systems.
\end{abstract}


%
\IEEEpeerreviewmaketitle


\section{Introduction}
\label{SectionI}
We study the algorithms that reduce the average complexity of maximum likelihood (ML) decoding of convolutional codes. By ML decoding, we mean the decoder uses code-search to find, and to guarantee the output of, the most likely codeword.

Forney showed that ML decoding of convolutional codes is equivalent to finding the most probable path taken through a Markov graph \cite{ref Forney73}. Denote the codeword length by $N$ and the coding memory by $\nu$. For each time index, the number of Markov states in the Markov graph is exponential in $\nu$. The total number of Markov states is therefore exponential in $\nu$ but linear in $N$. Define the complexity of a decoder as the number of visited Markov states normalized by the codeword length $N$. Practical ML decoding is often achieved using the Viterbi algorithm (VA) \cite{ref Viterbi67}\cite{ref Forney73}, whose complexity does not scale in $N$ but scales exponentially in $\nu$. Well known decoders such as the list decoders \cite{ref Zigangirov80}, the sequential decoders \cite{ref Fano63}, and the iterative decoders \cite{ref Bahl74} are able to achieve near optimal error performance with low average complexity. However, these decoders do not guarantee the output of the ML codeword \cite{ref Johannesson99}.

If obtaining the ML codeword is strictly enforced (see Section \ref{Discussions} for justification), to avoid exhaustive path search, the decoder must develop certain criterion or bound that can be used to disprove the optimality of a Markov path set. This is equivalent to developing an optimality test criterion (OTC) \cite{ref Swaszek98} to test whether the ML path (or codeword) belongs to the complementary path set (or codeword set)\footnote{In the literature such as \cite{ref Swaszek98}, OTC refers to a criterion designed to test whether a single codeword is optimum. In this paper, we extend the definition of OTC to a general criterion that can either verify or disprove the optimality of a codeword set.}.

Two major OTCs have been used in the ML decoding of convolutional codes. The first one is the ``path covering criterion" (PCC) (explained in \cite{ref Ariel99} and in Appendix \ref{PCC}) used in the VA \cite{ref Viterbi67}\cite{ref Forney73}. VA visits {\it all} Markov states in chronological order \cite{ref Forney73}. For each time index, the decoder maintains a set of ``cover" (defined in Appendix \ref{PCC}) Markov paths each passing one of the Markov states \cite{ref Forney73}. According to the PCC, the ``cover" Markov path passing a Markov state disproves the optimality of all other Markov paths passing the same state. The second OTC is the sum log likelihood (SLL)-based OTCs used extensively in the sphere decoder \cite{ref Fincke85}\cite{ref Hassibi05}. Sphere decoder models ML decoding as finding the lattice point closest to the channel output in the signal space \cite{ref Hassibi05}. Hence the distance between the channel output and an arbitrary lattice point upper bounds the distance from the channel output to the ML codeword. Such distance bound is based on the SLL of the corresponding codeword, and is used in the sphere decoder \cite{ref Fincke85}\cite{ref Hassibi05} as well as other ML decoders \cite{ref Swaszek98} as the key means to avoid exhaustive codeword search. In \cite{ref Vikalo02}\cite{ref Vikalo03}, Vikalo and Hassibi showed that PCC-based and SLL-based optimality tests can be combined to find the ML codeword without visiting all Markov states.

Assume PCC-based optimality test is always implemented. In this paper, we first show that {\it additional} complexity reduction brought by the SLL-based optimality test diminishes as one fixes the coding memory $\nu$ and takes the codeword length $N$ to infinity. Such inefficiency is due to the fact that SLL-based OTC does not exploit the structure of the convolutional code. Searching the ML codeword is equivalent to finding the ML source message, which contains a sequence of source symbols. We show whether the ML message contains a particular symbol at a given time index can be tested using an OTC that depends only on the log likelihood of channel output symbols in a {\it fixed-sized} time neighborhood. We call such test the neighboring log likelihood (NLL)-based optimality test, and show its efficiency does not depend on the codeword length. We theoretically demonstrate that NLL-based optimality test can bring significant complexity reduction to ML decoding when the communication system has a high signal to noise ratio (SNR). Complexity of the decoder using SLL-base optimality test, on the other hand, remains the same as the VA for all SNR if the codeword length is taken to infinity. The results are also generalized to ML sequence detection in a class of discrete-time hidden Markov systems \cite{ref Rabiner89}.

\section{Problem Formulation}
\label{SectionII}
Let $C$ be an $(n, k)$ convolutional code over $\mbox{GF}(q)$ defined by a polynomial generater matrix $\mbox{\boldmath  $G$}(D)$ \cite{ref Forney73c},
\begin{equation}
\mbox{\boldmath  $G$}(D)=\mbox{\boldmath  $G$}[0]+\mbox{\boldmath  $G$}[1]D+ \dots +\mbox{\boldmath  $G$}[\nu-1]D^{\nu-1},
\end{equation}
where $D$ is the delay operator; $\nu$ is the coding memory; $\mbox{\boldmath  $G$}[l]$, $\l=0,\dots,\nu-1$, are $k\times n$ matrices over $\mbox{GF}(q)$. Assume $\mbox{\boldmath  $G$}(D)$ is a minimal encoder \cite{ref Forney73c}.

Denote the source message by a {\it sequence} of vector {\it symbols},
\begin{equation}
\mbox{\boldmath  $x$}(D)=\mbox{\boldmath  $x$}[d]D^d+\mbox{\boldmath  $x$}[d+1]D^{d+1}+\dots,
\end{equation}
where $d$ is the time index, possibly negative; $\mbox{\boldmath  $x$}[d]$, $\forall d$, are row vectors of dimension $k$ over $\mbox{GF}(q)$. The encoded message, or the corresponding codeword, is given by
\begin{equation}
\mbox{\boldmath  $y$}(D)=\mbox{\boldmath  $x$}(D)\mbox{\boldmath  $G$}(D)=\sum_{d}\sum_{l=0}^{\nu-1}\mbox{\boldmath  $x$}[d-l]\mbox{\boldmath  $G$}[l]D^d.
\end{equation}
To simplify the presentation, we assume time index $d$ takes all integer values. We assume $\mbox{\boldmath  $x$}[d]=\mbox{\boldmath  $0$}$ for $d<0$ and $d\ge N$. We term $N$ the codeword length.

Define a function $g_q(y)$ that maps $y$ from $\mbox{GF}(q)$ to ${\cal R}$ (the set of real numbers) in one-to-one sense. If $\mbox{\boldmath  $y$}(D)$ is a vector sequence, $g_q(\mbox{\boldmath  $y$}(D))$ applies the mapping to each of the elements of $\mbox{\boldmath  $y$}(D)$, respectively\footnote{Hence the output of $g_q(\mbox{\boldmath  $y$}(D))$ is a vector sequence of the same length and dimension as $\mbox{\boldmath  $y$}(D)$.}. Assume the codeword is transmitted over a memoryless Gaussian channel. The channel output symbol sequence is given by
\begin{equation}
\mbox{\boldmath  $r$}(D)=g_q(\mbox{\boldmath  $y$}(D))+\mbox{\boldmath  $n$}(D)=g_q(\mbox{\boldmath  $x$}(D)\mbox{\boldmath  $G$}(D))+\mbox{\boldmath  $n$}(D),
\label{ChannelObservations}
\end{equation}
where $\mbox{\boldmath  $n$}(D)=\mbox{\boldmath  $n$}[d]D^d+\mbox{\boldmath  $n$}[d+1]D^{d+1}+\dots$ is the noise sequence with $\mbox{\boldmath  $n$}[d] \sim N(\mbox{\boldmath  $0$}, \sigma^2\mbox{\boldmath  $I$} )$ being i.i.d. Gaussian. Without loss of generality, we define the scaled signal to noise ratio of the system as $\mbox{SNR}=\frac{1}{\sigma^2}$. In Section \ref{HMM}, we show that the results are generalizable not only to other channel models, but also to a class of hidden Markov systems.

Given the channel output, for any source message $\mbox{\boldmath  $x$}(D)$ and its corresponding codeword $\mbox{\boldmath  $y$}(D)=\mbox{\boldmath  $x$}(D)\mbox{\boldmath  $G$}(D)$, we define the ``negative SLL" as
\begin{equation}
S_x(\mbox{\boldmath  $x$}(D))=S_y(\mbox{\boldmath  $y$}(D))=\sum_{d=0}^{N+\nu-1}\left\|\mbox{\boldmath  $r$}[d]-g_q\left(\mbox{\boldmath  $y$}[d]\right)\right\|^2.
\label{NegativeSLL}
\end{equation}
The objective of ML decoding is to find the ML message $\mbox{\boldmath  $x$}_{ML}(D)$ that minimizes the negative SLL,
\begin{equation}
\mbox{\boldmath  $x$}_{ML}(D)= \mathop{\mbox{argmin}}_{\mbox{\scriptsize \boldmath  $x$}[d], 0\le d<N}S_x(\mbox{\boldmath  $x$}(D)).
\end{equation}

Throughout this paper, we assume PCC-based optimality test is always implemented. For the sake of completeness, a description of PCC-based optimality test is given in Appendix \ref{PCC}.

\section{Inefficiency of Sum Log Likelihood-based Optimality Test}
\label{SectionIII}

For ML decoders using SLL-based optimality test, the decoder first obtains a quick guess of the source message without solving the ML decoding problem. SLL of the obtained message is then used to help disproving the optimality of certain Markov path sets and consequently to avoid exhaustive path search. We make an ideal assumption that the ``guessed" message equals the transmitted message\footnote{Note that the decoder still needs to testify whether the guessed message is indeed the ML solution. If it is not, then a search for the ML message must be carried out.}. We show in this section that, even under this ideal assumption, complexity reduction brought by the SLL-based optimality tests still diminishes as we take $N$ to infinity.

Let $\mbox{\boldmath  $x$}(D)$ be the actual source message, which is also the message ``guessed" by the decoder. Let $\mbox{\boldmath  $y$}(D)=\mbox{\boldmath  $x$}(D)\mbox{\boldmath  $G$}(D)$ be the transmitted codeword. The corresponding negative SLL is given by
\begin{equation}
S_x(\mbox{\boldmath  $x$}(D))=\sum_{d=0}^{N+\nu-1}\left\|\mbox{\boldmath  $r$}[d]-g_d\left(\mbox{\boldmath  $y$}[d]\right)\right\|^2=\sum_{d=0}^{N+\nu-1}\left\|\mbox{\boldmath  $n$}[d]\right\|^2.
\label{SLL}
\end{equation}

Now consider a subset of time indices $D_d^x\subseteq [0, N)$. Let $\left\{\tilde{\mbox{\boldmath  $x$}}[d]|d\in D_d^x \right\}$ be a {\it partial message} defined only at time indices in $D_d^x$. Denote by $\{\tilde{\mbox{\boldmath  $x$}}(D_d^x)\}$ the set of messages satisfying
\begin{equation}
\{\tilde{\mbox{\boldmath  $x$}}(D_d^x)\}=\{\mbox{\boldmath  $x$}_0(D)|\mbox{\boldmath  $x$}_0[d]=\tilde{\mbox{\boldmath  $x$}}[d], \forall d\in D_d^x, \mbox{\boldmath  $x$}_0(D)\ne \mbox{\boldmath  $x$}(D) \}.
\end{equation}
Suppose the decoder wants to test whether it can disprove the optimality of $\{\tilde{\mbox{\boldmath  $x$}}(D_d^x)\}$, i.e., whether $\mbox{\boldmath  $x$}_{ML}(D)\not\in\{\tilde{\mbox{\boldmath  $x$}}(D_d^x)\}$. A common practice \cite{ref Swaszek98}\cite{ref Vikalo02}\cite{ref Vikalo03} is to find a lower bound, denoted by $S_x^L(\tilde{\mbox{\boldmath  $x$}}(D_d^x))$, of the negative SLLs of the messages in $\{\tilde{\mbox{\boldmath  $x$}}(D_d^x)\}$.
\begin{equation}
S_x(\mbox{\boldmath  $x$}_0(D)) \ge S_x^L(\tilde{\mbox{\boldmath  $x$}}(D_d^x)), \qquad \forall \mbox{\boldmath  $x$}_0(D)\in \{\tilde{\mbox{\boldmath  $x$}}(D_d^x)\}.
\end{equation}
If the lower bound $S_x^L(\tilde{\mbox{\boldmath  $x$}}(D_d^x))$ is larger than $S_x(\mbox{\boldmath  $x$}(D))$ obtained in (\ref{SLL}), then we have $S_x(\mbox{\boldmath  $x$}_0(D))\ge S_x^L(\tilde{\mbox{\boldmath  $x$}}(D_d^x)) >S_x(\mbox{\boldmath  $x$}(D))$ for all $\mbox{\boldmath  $x$}_0(D)\in \{\tilde{\mbox{\boldmath  $x$}}(D_d^x)\}$, which means the ML message is not in $\{\tilde{\mbox{\boldmath  $x$}}(D_d^x)\}$.

In Appendix \ref{SNLLowerBound}, we show that the SLL lower bounds appeared in the literature satisfy the following assumption.
\begin{assumption}{\label{Assumption1}}
Given $\{\tilde{\mbox{\boldmath  $x$}}(D_d^x)\}$, let $D_d^y\subseteq [0, N+\nu)$ be the maximum time index set, over which we can find a {\it partial codeword} $\tilde{\mbox{\boldmath  $y$}}(D_d^y)$ such that for all $\mbox{\boldmath  $x$}_0(D)\in \{\tilde{\mbox{\boldmath  $x$}}(D_d^x)\}$ with $\mbox{\boldmath  $y$}_0(D)=\mbox{\boldmath  $x$}_0(D)\mbox{\boldmath  $G$}(D)$, we have $\mbox{\boldmath  $y$}_0[d]=\tilde{\mbox{\boldmath  $y$}}[d]$ for all $d\in D_d^y$. Note that $D_d^y$ and $\tilde{\mbox{\boldmath  $y$}}(D_d^y)$ are uniquely determined by $\{\tilde{\mbox{\boldmath  $x$}}(D_d^x)\}$. We also have $|D_d^y|\le |D_d^x|+\nu$.

We assume the existence of a positive constant $\epsilon \in (0, 1]$, whose value does {\it not} depend on $N$, such that
\begin{eqnarray}
&&S_x^L(\tilde{\mbox{\boldmath  $x$}}(D_d^x)) \le \sum_{d\in D_d^y}\left\|\mbox{\boldmath  $r$}[d]-g_q\left(\tilde{\mbox{\boldmath  $y$}}[d]\right)\right\|^2   +(N+\nu-|D_d^y|)(1-\epsilon)n\sigma^2.
\end{eqnarray}
$\QED$
\end{assumption}

As demonstrated in \cite{ref Vikalo02}\cite{ref Swaszek98}, if we fix $N$, using $S_x^L(\tilde{\mbox{\boldmath  $x$}}(D_d^x))>S_x(\mbox{\boldmath  $x$}(D))$ as the OTC to disprove the optimality of message set $\{\tilde{\mbox{\boldmath  $x$}}(D_d^x)\}$ can bring significant complexity reduction to ML decoding, especially under high SNR. However, if we define  $D_e \subseteq D_d^y$ as the subset of time indices corresponding to the erroneous codeword symbols, i.e.,
\begin{equation}
D_e=\{d| d\in D_d^y, \tilde{\mbox{\boldmath  $y$}}(d)\ne \mbox{\boldmath  $y$}(d)\},
\label{De}
\end{equation}
the following proposition shows that SLL-based optimality tests become inefficient if $N-|D_d^x|$ is taken to infinity while $|D_e|$ is kept finite.

\begin{lemma}{\label{Proposition1}}
Assume the generater matrix $\mbox{\boldmath  $G$}(D)$ is fixed, and therefore the constraint length $\nu$ is fixed. Consider message sets characterized by $\{\tilde{\mbox{\boldmath  $x$}}(D_d^x)\}$ for arbitrary $D_d^x$ but under the constraint of a fixed $D_e$, where $D_e\subseteq D_d^y$ is defined in (\ref{De}) and the derivation of $D_d^y$ is specified in Assumption \ref{Assumption1}.

If we fix SNR and take $N-|D_d^x|$ to infinity, we have
\begin{equation}
\lim_{N-|D_d^x|\to \infty}P\{S_x^L(\tilde{\mbox{\boldmath  $x$}}(D_d^x))>S_x(\mbox{\boldmath  $x$}(D))\}=0.
\label{IneffSLLOTC}
\end{equation}
If we first take $N-|D_d^x|$ to infinity and then take SNR to infinity, we have
\begin{equation}
\lim_{\mbox{\scriptsize SNR}\to \infty}\lim_{N-|D_d^x|\to \infty}P\{S_x^L(\tilde{\mbox{\boldmath  $x$}}(D_d^x))>S_x(\mbox{\boldmath  $x$}(D))\}=0.
\label{IneffSLLOTCSNR}
\end{equation}
$\QED$
\end{lemma}

\begin{proof}
Since $|D_d^y|\le |D_d^x|+\nu$, taking $N-|D_d^x|$ to infinity implies taking $N-|D_d^y|$ to infinity.

According to Assumption \ref{Assumption1}, we have
\begin{eqnarray}
\frac{S_x^L(\tilde{\mbox{\boldmath  $x$}}(D_d^x))-S_x(\mbox{\boldmath  $x$}(D))}{N+\nu-|D_d^y|}   & \le & \frac{1}{N+\nu-|D_d^y|}\left(\sum_{d\in D_e}\left\|\mbox{\boldmath  $r$}[d]-g_q\left(\tilde{\mbox{\boldmath  $y$}}[d]\right)\right\|^2 \right)   +(1-\epsilon)n\sigma^2  \nonumber \\
&& - \frac{1}{N+\nu-|D_d^y|}\left(\sum_{d\in D_e}\|\mbox{\boldmath  $n$}[d]\|^2 \right)  -\frac{1}{N+\nu-|D_d^y|}\sum_{d\not\in D_d^y}\|\mbox{\boldmath  $n$}[d]\|^2 .
\label{SLLBound2}
\end{eqnarray}

Since $\mbox{\boldmath  $n$}[d]$ are i.i.d. Gaussian with covariance matrix $\sigma^2\mbox{\boldmath  $I$}$, $\|\mbox{\boldmath  $n$}[d]\|^2$ are i.i.d. $\chi^2$ with mean $n\sigma^2$ and variance $2n\sigma^4$. Therefore $\frac{1}{N+\nu-|D_d^y|}\sum_{d\not\in D_d^y}\|\mbox{\boldmath  $n$}[d]\|^2\to n\sigma^2$, $\frac{1}{N+\nu-|D_d^y|}\left(\sum_{d\in D_e}\left\|\mbox{\boldmath  $r$}[d]-g_q\left(\tilde{\mbox{\boldmath  $y$}}[d]\right)\right\|^2 \right)\to 0$, and $\frac{1}{N+\nu-|D_d^y|}\sum_{d\in D_e}\|\mbox{\boldmath  $n$}[d]\|^2\to 0$ with probability one as $N-|D_d^y|\to \infty$. Consequently, denote the right hand side of (\ref{SLLBound2}) by $U_0$, we have with probability one,
\begin{equation}
\lim_{N-|D_d^y|\to \infty} U_0 = -\epsilon n\sigma^2 <0.
\end{equation}
This yields
\begin{eqnarray}
\lim_{N-|D_d^x|\to \infty}P\left\{S_x^L(\tilde{\mbox{\boldmath  $x$}}(D_d^x))>S_x(\mbox{\boldmath  $x$}(D))\right\}   &=& \lim_{N-|D_d^y|\to \infty}P\left\{\frac{S_x^L(\tilde{\mbox{\boldmath  $x$}}(D_d^x))-S_x(\mbox{\boldmath  $x$}(D))}{N+\nu-|D_d^y|} >0\right\}  \nonumber \\
&\le &  \lim_{N-|D_d^y|\to \infty}P\left\{U_0 >0\right\}=0.
\label{IneffSLLOTC}
\end{eqnarray}

Since (\ref{IneffSLLOTC}) holds for all SNR, the conclusion remains true if we take SNR to infinity after $N-|D_d^x|$ is taken to infinity\footnote{Note that the order in which limits are taken in (\ref{IneffSLLOTCSNR}) is important. If we fix $N$ and take SNR to infinity first, we can get $\lim_{N-|D_d^x|\to \infty}\lim_{\mbox{\tiny SNR}\to \infty}P\{S_x^L(\tilde{\mbox{\boldmath  $x$}}(D_d^x))>S_x(\mbox{\boldmath  $x$}(D))\}=1$.}.
\end{proof}

With the help of Lemma \ref{Proposition1}, inefficiency of SLL-based optimality tests is characterized by the following lemma.
\begin{lemma}{\label{Proposition2}}
Let $C_{sll}$ be the complexity of an ML decoder that only uses PCC- and SLL-based optimality tests for complexity reduction. Let $C_{va}$ be the complexity of the Viterbi decoder, in which, only PCC-based optimality test is used. For any $\delta>0$, we have,
\begin{eqnarray}
&& \lim_{N\to \infty} P\{C_{sll} \ge (1-\delta)C_{va}\}=1 \nonumber \\
&& \lim_{\mbox{\scriptsize SNR}\to \infty}\lim_{N\to \infty} P\{C_{sll} \ge (1-\delta)C_{va}\}=1.
\label{ComplexityLimit}
\end{eqnarray}
$\QED$
\end{lemma}

The proof of Lemma \ref{Proposition2} is given in Appendix \ref{ProofProposition2}.

\section{Neighboring Log Likelihood-based Optimality Test}
\label{SectionIV}

We propose in Theorem \ref{Theorem1} a class of NLL-based optimality tests, whose efficiency does not depend on the codeword length $N$. We show in Section \ref{SectionV} that these NLL-based optimality tests can significantly reduce the average complexity of ML decoding under high SNR. This is in contrast to the inefficiency of SLL-based optimality tests which are not able to bring meaningful complexity reduction if $N$ is taken to infinity first.

\begin{theorem}{\label{Theorem1}}
Define $d_{\min}^2$, $d_{\max}^2$ by
\begin{equation}
d_{\min}^2=\min_{\mbox{\scriptsize \boldmath  $y$}_1 \ne \mbox{\scriptsize \boldmath  $y$}_2}\|g_q(\mbox{\boldmath  $y$}_1)-g_q(\mbox{\boldmath  $y$}_2)\|^2, \qquad d_{\max}^2=\max_{\mbox{\scriptsize \boldmath  $y$}_1 \ne \mbox{\scriptsize \boldmath  $y$}_2}\|g_q(\mbox{\boldmath  $y$}_1)-g_q(\mbox{\boldmath  $y$}_2)\|^2,
\end{equation}
where $\mbox{\boldmath  $y$}_1$, $\mbox{\boldmath  $y$}_2$ are $n$-dimensional row vectors over $GF(q)$. Let $\xi$ be an arbitrary constant, $M$ be an arbitrary integer,  satisfying
\begin{equation}
0<\xi<\frac{d_{\min}^2}{2}, \qquad M>\frac{\nu d_{\max}^2}{3\xi}.
\label{xiMconstraint}
\end{equation}
Let $\mbox{\boldmath  $x$}_0(D)$ be a source message whose corresponding codeword is $\mbox{\boldmath  $y$}_0(D)$. For any time index $m$, if the following inequality is satisfied for all $d\in [m-2M\nu, m+2M\nu)$,
\begin{equation}
\|\mbox{\boldmath  $r$}[d]-g_q(\mbox{\boldmath  $y$}_0[d]) \| < \frac{d_{\min}^2}{2}-\xi,
\label{PLLDiference1}
\end{equation}
and the following inequalities hold,
\begin{eqnarray}
&& \sum_{d=m+2M\nu}^{m+(2M+1)\nu-1} \|\mbox{\boldmath  $r$}[d]-g_q(\mbox{\boldmath  $y$}_0[d]) \|^2 \le M\xi -\nu d_{\max}^2 \nonumber \\
&& \sum_{d=m-(2M+1)\nu}^{m-2M\nu-1} \|\mbox{\boldmath  $r$}[d]-g_q(\mbox{\boldmath  $y$}_0[d]) \|^2 \le M\xi -\nu d_{\max}^2,
\label{PLLDiference2}
\end{eqnarray}
then we must have $\mbox{\boldmath  $x$}_0[\tilde{m}]=\mbox{\boldmath  $x$}_{ML}[\tilde{m}]$, $\forall \tilde{m}\in [m, m+\nu)$. $\QED$
\end{theorem}

We skip the proof of Theorem \ref{Theorem1} since the result is implied by Theorem \ref{Theorem3} presented in Section \ref{HMM}.

Note that the values of $d_{\min}$ and $d_{\max}$ only depend on the $g_q()$ function. Hence, as long as $g_q()$ and $\nu$ are given, the values of $\xi$ and $M$ can be fixed, e.g., $\xi=\frac{d_{\min}^2}{4}$ and $M=\left\lceil\frac{4\nu d_{\max}^2}{3d_{\min}^2}\right\rceil$. Given $M$, the optimality test presented in Theorem \ref{Theorem1} testifies the optimality of $\{\mbox{\boldmath  $x$}[\tilde{m}]| \tilde{m}\in [m, m+\nu) \}$ using the log likelihood of channel output symbols within a {\it fixed-sized} time interval $[m-(2M+1)\nu, m+(2M+1)\nu)$. It is quite intuitive to see, efficiency of the test does not depend on the codeword length if all other parameters are fixed.

Efficiency of the OTC proposed in Theorem \ref{Theorem1} is characterized by the following lemma.

\begin{lemma}{\label{Proposition3}}
Assume $\xi$ and $M$ are chosen to satisfy (\ref{xiMconstraint}). Let $m$ be an arbitrary time index. Let $\mbox{\boldmath  $y$}_0(D)$ equal the transmitted codeword within time interval $[m-(2M+1)\nu, m+(2M+1)\nu)$. Define $\mbox{OPT}_m$ as the event that (\ref{PLLDiference2}) is satisfied and (\ref{PLLDiference1}) is satisfied for all $d\in [m-2M\nu, m+2M\nu)$.

Fix all other parameters and take SNR to infinity, we have
\begin{equation}
\lim_{\mbox{\scriptsize SNR}\to \infty} P\left\{\mbox{OPT}_m \right\}=1.
\label{PLLOTCEfficiency}
\end{equation}
The same conclusion holds if we first take $N$ to infinity, then take SNR to infinity.
\begin{equation}
\lim_{\mbox{\scriptsize SNR}\to \infty}\lim_{N\to \infty} P\left\{\mbox{OPT}_m \right\}=1.
\label{PLLOTCEfficiencyN}
\end{equation}
\end{lemma}

\begin{proof}
If $\mbox{\boldmath  $y$}_0(D)$ equals the transmitted codeword within time interval $[m-(2M+1)\nu, m+(2M+1)\nu)$, for $d\in[m-(2M+1)\nu, m+(2M+1)\nu)$, we have
\begin{equation}
\mbox{\boldmath  $r$}[d]-g_q(\mbox{\boldmath  $y$}_0[d])=\mbox{\boldmath  $n$}[d].
\end{equation}
Consequently, (\ref{PLLOTCEfficiency}) and (\ref{PLLOTCEfficiencyN}) hold because $\|\mbox{\boldmath  $n$}[d]\|^2$ are i.i.d. $\chi^2$, whose mean, $\frac{n}{\mbox{\scriptsize SNR}}$, and variance, $\frac{2n}{\mbox{\scriptsize SNR}^2}$, converge to $0$ as SNR goes to infinity.
\end{proof}

Lemma \ref{Proposition3} implies, if there is a suboptimal decoder whose probability of {\it symbol} detection error (as opposed to sequence detection error) is low under high SNR, then NLL-based optimality tests can help transforming the suboptimal detector to an ML detector with only marginal increase in average decoding complexity. An example of such transformation is presented in the following section.

\section{A Three-step ML Decoding Framework}
\label{SectionV}
The communication system given in Section \ref{SectionII} follows a discrete-time hidden Markov model \cite{ref Rabiner89}, where each Markov state at time index $d$ corresponds to a possible combination of source symbols in time interval $(d-\nu, d]$. If a decoder obtains the ML codeword using the VA, all Markov states within time interval $[\nu, N]$ have to be visited. Alternatively, if one can use a low complexity algorithm to disprove the optimality of most of the Markov states, then the VA can limit its search by visiting only a small subset of Markov states.

Following this idea, the three-step ML decoding framework is given as follows.

\begin{itemize}
\item Step 1: The decoder uses a suboptimal algorithm (denoted by $\Phi_{\mbox{\scriptsize sub}}$) to obtain a quick guess of the codeword $\tilde{\mbox{\boldmath  $y$}}(D)$ and its corresponding source message $\tilde{\mbox{\boldmath  $x$}}(D)$.

\item Step 2: An NLL-based optimality test (specified in Theorem \ref{Theorem1}) is applied to each of the source symbols of $\tilde{\mbox{\boldmath  $x$}}(D)$. The decoder maintains a source symbol set sequence $X(D)$, with $X[d]$ being the source symbol set of time index $d$. If $\tilde{\mbox{\boldmath  $x$}}[d]=\mbox{\boldmath  $x$}_{ML}[d]$ can be confirmed by the optimality test, we let $X[d]=\{\tilde{\mbox{\boldmath  $x$}}[d]\}$; otherwise, we let $X[d]$ be the set of all possible source symbol vectors at time index $d$.

\item Step 3: The decoder uses a modified VA to search for the ML source message. The only difference between the modified VA and the conventional VA is that, the modified VA visits a Markov state only if all source symbols corresponding to the Markov state belong to the source symbol sets $X[d]$ of the corresponding time indices.
\end{itemize}

Implementing the modified VA is quite straightforward. Hence its further description is skipped. Comparing to the three-step decoding algorithm studied in \cite{ref Swaszek98}, the key advantage of using an NLL-based optimality test is that the test can be applied to an individual source symbol rather than the whole source message.

\begin{theorem}{\label{Theorem2}}
Let $P_e\{\Phi_{\mbox{\scriptsize sub}}\}$ be the probability of {\it symbol} detection error of $\Phi_{\mbox{\scriptsize sub}}$. Assume, while fixing all other parameters,
\begin{equation}
\lim_{\mbox{\scriptsize SNR} \to \infty}P_e\{\Phi_{\mbox{\scriptsize sub}}\}=0, \qquad   \lim_{\mbox{\scriptsize SNR} \to \infty}\lim_{N\to \infty}P_e\{\Phi_{\mbox{\scriptsize sub}}\}=0.
\label{SymbolError}
\end{equation}
Let $C_{\mbox{\scriptsize mva}}$ be the average number of Markov states per time unit visited by the modified VA in the third step of the ML decoder. For any $\delta>0$, we have
\begin{equation}
\lim_{\mbox{\scriptsize SNR} \to \infty}P\{C_{\mbox{\scriptsize mva}}\le 1+\delta\}=1, \qquad   \lim_{\mbox{\scriptsize SNR} \to \infty}\lim_{N\to \infty}P\{C_{\mbox{\scriptsize mva}}\le 1+\delta\}=1.
\end{equation}
\end{theorem}

\begin{proof}
Let $\mbox{\boldmath  $x$}(D)$, $\mbox{\boldmath  $y$}(D)$ be the actual source message and the transmitted codeword, respectively. Let $\tilde{\mbox{\boldmath  $x$}}(D)$, $\tilde{\mbox{\boldmath  $y$}}(D)$ be the source message and the codeword output by $\Phi_{\mbox{\scriptsize sub}}$. According to (\ref{SymbolError}), for any time index $m$, we have
\begin{equation}
\lim_{\mbox{\scriptsize SNR} \to \infty}P\left\{\begin{array}{l}\tilde{\mbox{\boldmath  $y$}}[d]=\mbox{\boldmath  $y$}[d], \\ \forall d\in [m-2(M-1)\nu, m+(2M+1)\nu)\end{array}\right\}=1.
\label{CorrectDetection}
\end{equation}
where $M$ is the parameter of the NLL-based optimality test, specified in Theorem \ref{Theorem1}. According to (\ref{CorrectDetection}), Lemma \ref{Proposition2}, and Theorem \ref{Theorem1}, for any $m$, if $\tilde{\mbox{\boldmath  $y$}}[d]=\mbox{\boldmath  $y$}[d], \forall d\in [m-2(M-1)\nu, m+(2M+1)\nu)$, then the probability that the NLL-based optimality test can confirm $\tilde{\mbox{\boldmath  $x$}}[d]=\mbox{\boldmath  $x$}_{ML}[d], \forall d\in [m, m+\nu)$ converges to one as $\mbox{SNR} \to \infty$. Consequently, letting $X[d]$ be the source symbol set maintained by the ML decoder in the second step, we have
\begin{equation}
\lim_{\mbox{\scriptsize SNR} \to \infty}P\left\{|X[d]|=1, \forall d\in [m, m+\nu)\right\}=1, \qquad \forall m
\label{LowComplexityinProb}
\end{equation}
Since the worst case complexity of the modified VA is bounded, (\ref{LowComplexityinProb}) implies, for any $\delta>0$, $\lim_{\mbox{\scriptsize SNR} \to \infty}P\{C_{\mbox{\scriptsize mva}}\le 1+\delta\}=1$.

Since all derivations hold if we first take $N$ to infinity, we also have $ \lim_{\mbox{\scriptsize SNR} \to \infty}\lim_{N\to \infty}P\{C_{\mbox{\scriptsize mva}}\le 1+\delta\}=1$.
\end{proof}

By sharing computations among optimality tests, it is easy to see that the complexity of the second step of the ML decoder is equivalent, in order, to visiting one Markov state per time unit. Therefore, if $\Phi_{\mbox{\scriptsize sub}}$ satisfies (\ref{SymbolError}), as $\mbox{SNR} \to \infty$, the complexity of the three-step ML decoder converges to the complexity of $\Phi_{\mbox{\scriptsize sub}}$, which can be significantly lower than the complexity of the VA. Moreover, the three steps of the ML decoder can be implemented in a parallelized manner in the sense that each step can process some of the source symbols without waiting for the previous step to {\it completely} finish its work. An example of such parallelized implementation can be found in \cite[The Simple MLSD Algorithm]{ref Luo07}.

\section{Maximum Likelihood Sequence Detection in A Class of Hidden Markov Systems}
\label{HMM}

In this section, we generalize the results of Section \ref{SectionIV} to ML sequence detection (MLSD) in a class of first order discrete-time hidden Markov systems \cite{ref Rabiner89}. We demonstrate in Appendix \ref{HMMModelVerification} that the communication system presented in Section \ref{SectionII} satisfies the model and the key assumptions given in this section.

Let $\mbox{\boldmath  $u$}(D)=\mbox{\boldmath  $u$}[d]D^d+\mbox{\boldmath  $u$}[d+1]D^{d+1}+...$ be a first order Markov sequence, where $d$ is the time index, possibly negative; $\mbox{\boldmath  $u$}[d]$ represents the Markov state (at time $d$), which is a $k_{\nu}$-dimensional row vector defined over $GF(q)$. We assume $\mbox{\boldmath  $u$}[d]=\mbox{\boldmath  $0$}$ for $d<0$ and $d\ge N$, with $N$ being the sequence length. Define $\mbox{\boldmath  $y$}[d]=\mbox{\boldmath  $y$}(\mbox{\boldmath  $u$}[d])$ as the ``processed state", which is a {\it deterministic} function of $\mbox{\boldmath  $u$}[d]$. $\mbox{\boldmath  $y$}[d]$ is a $n$-dimensional row vector defined over $GF(q)$. We term $\mbox{\boldmath  $y$}(D)=\mbox{\boldmath  $y$}[d]D^d+\mbox{\boldmath  $y$}[d+1]D^{d+1}+...$ the processed state sequence. Let $\mbox{\boldmath  $r$}(D)=\mbox{\boldmath  $r$}[d]D^d+\mbox{\boldmath  $r$}[d+1]D^{d+1}+...$ be the observation sequence, where $\mbox{\boldmath  $r$}[d]$ is a $n$-dimensional row vector with real-valued elements.

Denote the state transition probability of the hidden Markov system by
\begin{equation}
P_t(\mbox{\boldmath  $u$}_1|\mbox{\boldmath  $u$}_2)=P\{\mbox{\boldmath  $u$}[d+1]=\mbox{\boldmath  $u$}_1|\mbox{\boldmath  $u$}[d]=\mbox{\boldmath  $u$}_2\}.
\end{equation}
Define the transition probability ratio bound $p_{tr}$ by
\begin{equation}
p_{tr}=\min_{\scriptsize \begin{array}{c} \mbox{\scriptsize \boldmath  $u$}_1, \mbox{\scriptsize \boldmath  $u$}_2, P_t(\mbox{\scriptsize \boldmath  $u$}_1|\mbox{\scriptsize \boldmath  $u$}_2)>0 \\ \mbox{\scriptsize \boldmath  $u$}_3, \mbox{\scriptsize \boldmath  $u$}_4, P_t(\mbox{\scriptsize \boldmath  $u$}_3|\mbox{\scriptsize \boldmath  $u$}_4)>0 \end{array} } \frac{P_t(\mbox{\boldmath  $u$}_1|\mbox{\boldmath  $u$}_2)}{P_t(\mbox{\boldmath  $u$}_3|\mbox{\boldmath  $u$}_4)}.
\label{PTR}
\end{equation}
We assume the Markov chain is ergodic and homogeneous. Therefore, there exists a positive integer $\nu$, such that
\begin{equation}
P\{\mbox{\boldmath  $u$}[d+\nu]=\mbox{\boldmath  $u$}_1| \mbox{\boldmath  $u$}[d]=\mbox{\boldmath  $u$}_2\} \ne 0, \quad \forall \mbox{\boldmath  $u$}_1, \mbox{\boldmath  $u$}_2.
\label{Homogeneous}
\end{equation}

Denote the observation distribution function by
\begin{equation}
F_o(\mbox{\boldmath  $r$}|\mbox{\boldmath  $y$}_1)=P\{\mbox{\boldmath  $r$}[d]\le \mbox{\boldmath  $r$}|\mbox{\boldmath  $y$}[d]=\mbox{\boldmath  $y$}_1\}.
\label{ObservationFunction}
\end{equation}
Let the corresponding probability density function (or probability mass function) be $f_o(\mbox{\boldmath  $r$}|\mbox{\boldmath  $y$}_1)$.

We also make the following two key assumptions.
\begin{assumption}{\label{Assumption2}}
We assume state processing $\mbox{\boldmath  $y$}[d]=\mbox{\boldmath  $y$}(\mbox{\boldmath  $u$}[d])$ does not compromise the observability of the Markov states in the sense that there exists a positive integer $\nu$ satisfying the following property. Given two Markov state sequences $\mbox{\boldmath  $u$}(D)$ and $\tilde{\mbox{\boldmath  $u$}}(D)$. For any time index $d$, if $\mbox{\boldmath  $u$}[d]\ne \tilde{\mbox{\boldmath  $u$}}[d]$, then we can find a time index $m\in (d-\nu, d+\nu)$, such that $\mbox{\boldmath  $y$}(\mbox{\boldmath  $u$}[m])\ne \mbox{\boldmath  $y$}(\tilde{\mbox{\boldmath  $u$}}[m])$.
\end{assumption}

Note that we used the same constant $\nu$ in (\ref{Homogeneous}) and in Assumption \ref{Assumption2}. This is valid because if (\ref{Homogeneous}) is satisfied for $\nu=\nu_0$, then it is also satisfied for all $\nu \ge \nu_0$; similar property applies to Assumption \ref{Assumption2}. Consequently, if Assumption \ref{Assumption2} holds, a common integer $\nu$ satisfying both (\ref{Homogeneous}) and Assumption \ref{Assumption2} can always be found.

\begin{assumption}{\label{Assumption3}}
Assume the existence of two functions: $L_l(\mbox{\boldmath  $r$}, \mbox{\boldmath  $y$}_1)$ and $L_u(\mbox{\boldmath  $r$}, \mbox{\boldmath  $y$}_1)$, both are functions of the channel output symbol $\mbox{\boldmath  $r$}$ and the processed state $\mbox{\boldmath  $y$}_1$. Assume $L_l(\mbox{\boldmath  $r$}, \mbox{\boldmath  $y$}_1)$ and $L_u(\mbox{\boldmath  $r$}, \mbox{\boldmath  $y$}_1)$ have the following two properties.

First, the following inequalities hold for all $\mbox{\boldmath  $r$}$ and $\mbox{\boldmath  $y$}_1$.
\begin{eqnarray}
&& L_l(\mbox{\boldmath  $r$}, \mbox{\boldmath  $y$}_1) \le \min_{\mbox{\scriptsize \boldmath  $y$}_2, \mbox{\scriptsize \boldmath  $y$}_2\ne \mbox{\scriptsize \boldmath  $y$}_1}\left[-\log(f_o(\mbox{\boldmath  $r$}|\mbox{\boldmath  $y$}_2))+\log(f_o(\mbox{\boldmath  $r$}|\mbox{\boldmath  $y$}_1))\right] \nonumber \\
&& L_u(\mbox{\boldmath  $r$}, \mbox{\boldmath  $y$}_1) \ge \max_{\mbox{\scriptsize \boldmath  $y$}_2\ne \mbox{\scriptsize \boldmath  $y$}_3}\left[-\log(f_o(\mbox{\boldmath  $r$}|\mbox{\boldmath  $y$}_2))+\log(f_o(\mbox{\boldmath  $r$}|\mbox{\boldmath  $y$}_3))\right].
\label{NLLu}
\end{eqnarray}

Second, the complexity of evaluating $L_l(\mbox{\boldmath  $r$}, \mbox{\boldmath  $y$}_1)$ and $L_u(\mbox{\boldmath  $r$}, \mbox{\boldmath  $y$}_1)$ is low in the sense that they do not require the search of any processed state other than $\mbox{\boldmath  $y$}_1$. $\QED$
\end{assumption}

Note that validity of the results presented in this section does not depend on the second property imposed in Assumption \ref{Assumption3}. However, we still include the property in the assumption since the key motivation of posing Assumption \ref{Assumption3} is to use the two functions $L_l(\mbox{\boldmath  $r$}, \mbox{\boldmath  $y$}_1)$ and $L_u(\mbox{\boldmath  $r$}, \mbox{\boldmath  $y$}_1)$ as tools to avoid exhaustive Markov state search and hence to reduce the complexity of ML decoding. Also note that the right hand side of the second inequality in (\ref{NLLu}) is not a function of $\mbox{\boldmath  $y$}_1$. However, the upper bound on the left hand side is a function of a processed state $\mbox{\boldmath  $y$}_1$ since one often needs a ``reference state" in order to upper bound the right hand side of (\ref{NLLu}). Further explanation is given in Appendix \ref{HMMModelVerification}.

Given the observation sequence $\mbox{\boldmath  $r$}(D)$, the negative SLL of a state sequence $\mbox{\boldmath  $u$}(D)$ is obtained by
\begin{equation}
S_u(\mbox{\boldmath  $u$}(D))=-\sum_{d=0}^{N}\log(f_o(\mbox{\boldmath  $r$}[d]|\mbox{\boldmath  $y$}[d])P_t(\mbox{\boldmath  $u$}[d]|\mbox{\boldmath  $u$}[d-1])).
\end{equation}
The objective of MLSD is to find the ML sequence that minimizes the negative SLL,
\begin{equation}
\mbox{\boldmath  $u$}_{ML}(D)=\mathop{\mbox{argmin}}_{\mbox{\scriptsize \boldmath  $u$}[d], 0\le d<N}S_u(\mbox{\boldmath  $u$}(D)).
\end{equation}

The following theorem gives a class of NLL-based optimality tests.
\begin{theorem}{\label{Theorem3}}
Assume the discrete-time Markov system satisfies Assumptions \ref{Assumption2} and \ref{Assumption3}.

Let $\rho>0$ be a positive constant. Given a Markov state sequence $\mbox{\boldmath  $u$}(D)$ and the corresponding processed states $\mbox{\boldmath  $y$}(D)$. Let $p_{tr}$ be defined by (\ref{PTR}). For any time index $m$, if there is an integer $M>0$ such that for all $d\in [m-2M\nu, m+2M\nu)$
\begin{equation}
L_l(\mbox{\boldmath  $r$}[d], \mbox{\boldmath  $y$}[d]) > 3\nu(\rho-\log p_{tr}),
\label{LBound}
\end{equation}
and
\begin{eqnarray}
&& \sum_{d=m+2M\nu}^{m+(2M+1)\nu-1}L_u(\mbox{\boldmath  $r$}, \mbox{\boldmath  $y$}[d]) \le 3M\nu\rho + (\nu+1)\log p_{tr} \nonumber \\
&& \sum_{d=m-(2M+1)\nu}^{m-2M\nu-1}L_u(\mbox{\boldmath  $r$}, \mbox{\boldmath  $y$}[d]) \le 3M\nu\rho + \nu\log p_{tr},
\label{LMBound}
\end{eqnarray}
then $\mbox{\boldmath  $u$}[m+\nu-1]=\mbox{\boldmath  $u$}_{ML}[m+\nu-1]$ must be true. $\QED$
\end{theorem}

The proof of Theorem \ref{Theorem3} is given in Appendix \ref{ProofTheorem3}. Note that Theorem \ref{Theorem3} implies Theorem \ref{Theorem1} if we set the parameters in Theorem \ref{Theorem1} at the corresponding values given in Appendix \ref{HMMModelVerification}.

For communication systems following a discrete-time hidden Markov model, $f_o(\mbox{\boldmath  $r$}|\mbox{\boldmath  $y$}_1)$ often belongs to an ensemble of density (or probability) functions, with the actual realization determined by the SNR. In other words, we can write the observation density (or probability) $f_o(\mbox{\boldmath  $r$}|\mbox{\boldmath  $y$}_1, \mbox{SNR})$ as a function of the SNR. Assume the discrete-time Markov system satisfies Assumption \ref{Assumption3}, where both functions $L_l(\mbox{\boldmath  $r$}, \mbox{\boldmath  $y$}_1)$ and $L_u(\mbox{\boldmath  $r$}, \mbox{\boldmath  $y$}_1)$ can be functions of the SNR. We make the following assumption.

\begin{assumption}{\label{Assumption4}}
Assume the observation density (or probability) $f_o(\mbox{\boldmath  $r$}|\mbox{\boldmath  $y$}_1, \mbox{SNR})$ is a function of the SNR. Assume the discrete-time Markov system satisfies Assumption \ref{Assumption3}. Let the actual state sequence and the processed state sequence be $\mbox{\boldmath  $u$}(D)$ and $\mbox{\boldmath  $y$}(D)$, respectively. Define two positive numbers $d_{\min}^2$ and $d_{\max}^2$ as follows
\begin{eqnarray}
&& \frac{d_{\min}^2}{2}=  \sup\left\{\gamma\ge 0; \lim_{\mbox{\scriptsize SNR}\to \infty}P\{L_l(\mbox{\boldmath  $r$}[d], \mbox{\boldmath  $y$}[d])\ge \gamma \mbox{SNR}\}=1 \right\}, \nonumber \\
&& d_{\max}^2=    \inf\left\{\gamma\ge 0; \lim_{\mbox{\scriptsize SNR}\to \infty}P\{L_u(\mbox{\boldmath  $r$}[d], \mbox{\boldmath  $y$}[d])\le \gamma \mbox{SNR}\}=1 \right\}.
\end{eqnarray}
We assume
\begin{equation}
d_{\min}^2>0, \qquad d_{\max}^2<\infty.
\end{equation}
\end{assumption}

The following lemma characterizes the efficiency of the OTC proposed in Theorem \ref{Theorem3}.

\begin{lemma}{\label{Proposition4}}
Assume the discrete-time Markov system satisfies Assumptions \ref{Assumption2} and \ref{Assumption4}. Let the state sequence be $\mbox{\boldmath  $u$}(D)$. Let $\xi$ be an arbitrary constant, $M$ be an arbitrary integer, satisfying
\begin{equation}
0<\xi< \frac{d_{\min}^2}{2}, \qquad M> \frac{\nu d_{\max}^2}{\xi}.
\end{equation}
Let $\rho=\frac{\xi\mbox{\scriptsize SNR}}{3\nu}$. Given an arbitrary time index $m$, define $\mbox{OPT}_m$ as the event that (\ref{LMBound}) is satisfied and (\ref{LBound}) is satisfied for all $d\in [m-2M\nu, m+2M\nu)$. If we fix all other parameters except the SNR, we have
\begin{equation}
\lim_{\mbox{\scriptsize SNR}\to \infty}P\{\mbox{OPT}_m\}=1.
\end{equation}
If we fix all other parameters except the SNR and the sequence length $N$, we have
\begin{equation}
\lim_{\mbox{\scriptsize SNR}\to \infty}\lim_{N\to \infty}P\{\mbox{OPT}_m\}=1.
\end{equation}
\end{lemma}

We skip the proof of Lemma \ref{Proposition4} since it is quite straightforward.

Note that in Lemma \ref{Proposition4}, when we take $N$ and SNR to infinity, $M$ can be fixed at a constant. This indicates that, when testing the optimality of a Markov state at a given time index, the NLL-based optimality test only uses observation symbols in a fixed-sized time neighborhood. Based on Theorem \ref{Theorem3} and Lemma \ref{Proposition3}, a three-step ML sequence detector similar to the one presented in Section \ref{SectionV} can be developed to transform a suboptimal sequence detector to a low complexity ML sequence detector. The detailed discussion is skipped since it does not essentially differ from the one presented in Section \ref{SectionV}.

\section{Further Discussions}
\label{Discussions}
In a practical system, suboptimal decoders such as the belief-propagation-based iterative decoders \cite{ref Bahl74}\cite{ref Johannesson99} can achieve near optimal error performance with low complexity. It is natural to ask: if suboptimal decoding only causes a negligible performance loss, why one should even bother with enforcing the ML solution? Note that this question does not suggest a default answer since the argument can also be presented in the opposite direction, i.e., if ML decoding only causes a negligible complexity increase, why one should not use an ML decoder? Nevertheless, the purpose of our work is not to participate in the debate whether ML decoding is practically useful. Rather, one should interpret Theorem \ref{Theorem2} as, for convolutional codes, the existence of a well-performed low complexity suboptimal algorithm implies that ML decoding can be carried out with a similar complexity under high SNR. More importantly, such conclusion holds irrespective of the codeword length.

Although the efficiency of SLL-based optimality tests does not depend on the codeword length, NLL-based optimality tests are inefficient only when the codeword length is large. Lemma \ref{Proposition1} and Theorem \ref{Theorem2} suggest that complexity reduction brought by NLL-based optimality tests can be superior to SLL-based optimality tests even for moderate SNR if the codeword length is large enough.

\appendix
\subsection{The Path Covering Criterion}
\label{PCC}
Assume the discrete-time hidden Markov model given in Section \ref{HMM}\footnote{It is shown in Appendix \ref{HMMModelVerification} that the model is satisfied by the communication system given in Section \ref{SectionII}.}. Given the observation sequence $\mbox{\boldmath  $r$}(D)$. Let $\tilde{\mbox{\boldmath  $u$}}(D)$ and $\mbox{\boldmath  $u$}(D)$ be two Markov state sequences whose corresponding processed state sequences are $\tilde{\mbox{\boldmath  $y$}}(D)$ and $\mbox{\boldmath  $y$}(D)$, respectively. If we can find two time indices $d_1<d_2$, such that $\tilde{\mbox{\boldmath  $u$}}[d_1]=\mbox{\boldmath  $u$}[d_1]$, $\tilde{\mbox{\boldmath  $u$}}[d_2]=\mbox{\boldmath  $u$}[d_2]$, and
\begin{equation}
\sum_{d=d_1+1}^{d_2}\log\frac{f_o(\mbox{\boldmath  $r$}[d]|\tilde{\mbox{\boldmath  $y$}}[d-1])P_t(\tilde{\mbox{\boldmath  $u$}}[d]|\tilde{\mbox{\boldmath  $u$}}[d-1])}{f_o(\mbox{\boldmath  $r$}[d]|\mbox{\boldmath  $y$}[d-1]))P_t(\mbox{\boldmath  $u$}[d]|\mbox{\boldmath  $u$}[d-1])} < 0,
\label{PathCovering}
\end{equation}
we say $\mbox{\boldmath  $u$}(D)$ ``covers" $\tilde{\mbox{\boldmath  $u$}}(D)$.

{\bf Path Covering Criterion: }  Markov state sequence $\tilde{\mbox{\boldmath  $u$}}(D)$ cannot be the ML sequence if we can find another state sequence $\mbox{\boldmath  $u$}(D)$ that covers $\tilde{\mbox{\boldmath  $u$}}(D)$.

The proof of the PCC is skipped since it is quite well known \cite{ref Ariel99}.

We say $\mbox{\boldmath  $u$}(D)$ is a ``cover" path with respect to Markov states $\mbox{\boldmath  $u$}[d_1]$ and $\mbox{\boldmath  $u$}[d_2]$ at time indices $d_1<d_2$ if, among all Markov paths passing $\mbox{\boldmath  $u$}[d_1]$ and $\mbox{\boldmath  $u$}[d_2]$, $\mbox{\boldmath  $u$}(D)$ maximizes $\sum_{d=d_1+1}^{d_2}\log(f_o(\mbox{\boldmath  $r$}[d]|\mbox{\boldmath  $y$}[d-1])P_t(\mbox{\boldmath  $u$}[d]|\mbox{\boldmath  $u$}[d-1]))$. Assume all Markov paths start from $\mbox{\boldmath  $u$}[-1]=\mbox{\boldmath  $0$}$. We say $\mbox{\boldmath  $u$}(D)$ is a ``cover" path with respect to Markov state  $\mbox{\boldmath  $u$}[d_1]$ at time index $d_1>0$ if, among all Markov paths passing $\mbox{\boldmath  $u$}[d_1]$, $\mbox{\boldmath  $u$}(D)$ maximizes $\sum_{d=1}^{d_1}\log(f_o(\mbox{\boldmath  $r$}[d]|\mbox{\boldmath  $y$}[d-1])P_t(\mbox{\boldmath  $u$}[d]|\mbox{\boldmath  $u$}[d-1]))$.

\subsection{Examples of SLL-based Optimality Tests Satisfying Assumption \ref{Assumption1}}
\label{SNLLowerBound}

In \cite{ref Vikalo03}\cite{ref Vikalo02}, when the decoder branches a Markov path at time index $m<N$, the branch is characterized by a {\it partial message} $\{\tilde{\mbox{\boldmath  $x$}}[0], \tilde{\mbox{\boldmath  $x$}}[1], \dots, \tilde{\mbox{\boldmath  $x$}}[m]\}$. For any codeword $\tilde{\mbox{\boldmath  $y$}}(D)$ associated to the branch, we have
\begin{equation}
\tilde{\mbox{\boldmath  $y$}}[d]=\sum_{l=0}^{\nu-1}\tilde{\mbox{\boldmath  $x$}}[d-l]\mbox{\boldmath  $G$}[l].
\end{equation}
In other words, $D_d=[0, m]$. The negative SLL lower bound is given by
\begin{eqnarray}
&& S_y(\tilde{\mbox{\boldmath  $y$}}(D))=\sum_{d=0}^{N+\nu-1}\left\|\mbox{\boldmath  $r$}[d]-g_q\left(\tilde{\mbox{\boldmath  $y$}}[d]\right)\right\|^2    \ge \sum_{d=0}^{m}\left\|\mbox{\boldmath  $r$}[d]-g_q\left(\sum_{l=0}^{\nu-1}\tilde{\mbox{\boldmath  $x$}}[d-l]\mbox{\boldmath  $G$}[l]\right)\right\|^2,
\label{SphereBound}
\end{eqnarray}
which satisfies Assumption \ref{Assumption1} with $\epsilon=1$.

In \cite{ref Swaszek98}, several SLL-based OTCs were presented for decoding block codes. The decoder obtains a first guess $\mbox{\boldmath  $y$}(D)$ of the codeword. A negative SLL lower bound $S_y^L\le S_y(\tilde{\mbox{\boldmath  $y$}}(D)\ne \mbox{\boldmath  $y$}(D))$ is then developed for the codeword set $\{\tilde{\mbox{\boldmath  $y$}}(D)\ne \mbox{\boldmath  $y$}(D) \}$, which corresponds to the case of $D_d$ being an empty set in the context of Section \ref{SectionIII}. $\mbox{\boldmath  $y$}(D)$ is optimal if the optimality test $S_y^L > S_y(\mbox{\boldmath  $y$}(D))$ gives a positive answer \cite{ref Swaszek98}.

The lower bounds $S_y^L$  presented in \cite[Section III]{ref Swaszek98} satisfy the following inequality,
\begin{equation}
S_y^L \le \min_{\tilde{\mbox{\scriptsize \boldmath  $y$}}(D)\ne \mbox{\scriptsize \boldmath  $y$}(D)}\sum_{d=0}^{N+\nu-1}\left\|g_q\left(\tilde{\mbox{\boldmath  $y$}}[d]\right)-g_q\left(\mbox{\boldmath  $y$}[d]\right)\right\|^2
\label{DistanceBound}
\end{equation}
Since the coding constraint is $\nu$, we can always find a codeword $\tilde{\mbox{\boldmath  $y$}}(D)\ne \mbox{\boldmath  $y$}(D)$ with $\tilde{\mbox{\boldmath  $y$}}(D)$ differing from $\mbox{\boldmath  $y$}(D)$ at no more than $\nu$ codeword symbols. This implies that the right hand side of (\ref{DistanceBound}) can be upper bounded by a constant, denoted by $U_1$, which is not a function of $N$.
\begin{equation}
S_y^L \le \min_{\tilde{\mbox{\scriptsize \boldmath  $y$}}(D)\ne \mbox{\scriptsize \boldmath  $y$}(D)}\sum_{d=0}^{N+\nu-1}\left\|g_q\left(\tilde{\mbox{\boldmath  $y$}}[d]\right)-g_q\left(\mbox{\boldmath  $y$}[d]\right)\right\|^2\le U_1
\end{equation}
Consequently, given $\mbox{SNR}>0$ and $0<\epsilon<1$, there exists a constant $N_0$ such that Assumption \ref{Assumption1} is satisfied for $N>N_0$.

\subsection{Proof of Lemma \ref{Proposition2}}
\label{ProofProposition2}
\begin{proof}
Assume, in searching the ML codeword, the decoder successfully avoided visiting a Markov state specified by $\{\mbox{\boldmath  $x$}_0[d-\nu+1], \dots, \mbox{\boldmath  $x$}_0[d]\}$. This implies that we can find two time index sets, $D_0^x \subset [d-\nu+1, d]$ and $D_d^x$,  $D_d^x\cap [d-\nu+1, d]=\phi$, such that the optimality of {\it all} message sets $\{\tilde{\mbox{\boldmath  $x$}}(D_0^x\cup D_d^x)\}$ with $\tilde{\mbox{\boldmath  $x$}}[\tilde{d}]=\mbox{\boldmath  $x$}_0[\tilde{d}]$, $\forall \tilde{d}\in D_0^x$ is disproved. We choose $D_d^x$ with {\it the maximum cardinality} while make sure that, in disproving the optimality of $\{\mbox{\boldmath  $x$}_0[d-\nu+1], \dots, \mbox{\boldmath  $x$}_0[d]\}$, the detector visited {\it all} the Markov states $\{\tilde{\mbox{\boldmath  $x$}}[\tilde{d}-\nu+1], \dots, \tilde{\mbox{\boldmath  $x$}}[\tilde{d}]\}$ satisfying $[\tilde{d}-\nu+1, \tilde{d}]\subseteq D_d^x$.

According to the definitions of $D_0^x$ and $D_d^x$, the decoder needs to disprove the optimality of a special message set $\{\mbox{\boldmath  $x$}_0(D_0^x\cup D_d^x)\}$ defined by $\mbox{\boldmath  $x$}_0[\tilde{d}]=\mbox{\boldmath  $x$}_0[\tilde{d}]$, $\forall \tilde{d}\in D_0^x$ and $\mbox{\boldmath  $x$}_0[\tilde{d}]=\mbox{\boldmath  $x$}[\tilde{d}]$, $\forall \tilde{d}\in D_d^x$. The definition of $D_d^x$ also implies that the decoder needs to obtain a lower bound $S_x^L(\tilde{\mbox{\boldmath  $x$}}(D_0^x\cup D_d^x))$ of the negative SLLs of the messages in $\{\tilde{\mbox{\boldmath  $x$}}(D_0^x\cup D_d^x)\}$. The lower bound $S_x^L(\tilde{\mbox{\boldmath  $x$}}(D_0^x\cup D_d^x))$ should only be a function of the partial message $\tilde{\mbox{\boldmath  $x$}}(D_0^x\cup D_d^x)$, but should not depend on any source message symbol whose time index is outside $D_0^x\cup D_d^x$. However, since the corresponding $D_e$ (defined in (\ref{De})) of $\{\mbox{\boldmath  $x$}_0(D_0^x\cup D_d^x)\}$ satisfies $|D_e|\le 2\nu$, according to Lemma \ref{Proposition1}, the probability of disproving the optimality of $\{\mbox{\boldmath  $x$}_0(D_0^x\cup D_d^x)\}$ (using SLL-based optimality test) is low if $N-|D_0^x \cup D_d^x| \gg 2\nu$.

To make the argument explicit, the fact that the decoder visits {\it all} Markov states $\{\tilde{\mbox{\boldmath  $x$}}[\tilde{d}-\nu+1], \dots, \tilde{\mbox{\boldmath  $x$}}[\tilde{d}]\}$ with $[\tilde{d}-\nu+1, \tilde{d}]\subseteq D_d^x$ implies
\begin{equation}
C_{sll} \ge \frac{|D_d^x|-\nu}{N+\nu}C_{va}.
\label{ComplexityVA}
\end{equation}
According to Lemma \ref{Proposition1}, for any positive constant $\delta>0$, if we fix all other parameters and take $N$ to infinity, we have\footnote{An equivalent statement of (\ref{ComplexityProb2}) is, if $\frac{N-|D_d^x|-|D_0^x|}{2\nu}<\frac{\delta}{2\nu}N$, as $N\to \infty$, the probability of disproving the optimality of all message sets $\{\tilde{\mbox{\boldmath  $x$}}(D_0^x\cup D_d^x)\}$ with $\tilde{\mbox{\boldmath  $x$}}[\tilde{d}]=\mbox{\boldmath  $x$}_0[\tilde{d}]$, $\forall \tilde{d}\in D_0^x$, using SLL-based optimality test goes to zero.}
\begin{equation}
\lim_{N\to \infty}P\left\{ \frac{N-|D_d^x|-|D_0^x|}{2\nu}<\frac{\delta}{2\nu}N \right\}=1.
\label{ComplexityProb2}
\end{equation}
Combining (\ref{ComplexityVA}) and (\ref{ComplexityProb2}), we get
\begin{equation}
\lim_{N\to \infty}P\left\{C_{sll} \ge (1-\delta)C_{va} \right\}=1.
\label{ComplexityResult}
\end{equation}

Since (\ref{ComplexityResult}) holds for any fixed SNR, it still holds if we take SNR to infinity after taking $N$ to infinity, i.e.,
\begin{equation}
\lim_{\mbox{\scriptsize SNR}\to \infty}\lim_{N\to \infty}P\left\{C_{sll} \ge (1-\delta)C_{va} \right\}=1.
\end{equation}
\end{proof}

\subsection{The Hidden Markov Model and Its Key Assumptions}
\label{HMMModelVerification}

In this section, we show the communication system presented in Section \ref{SectionII} satisfies the discrete-time hidden Markov model and the key assumptions given in Section \ref{HMM}.

Consider a communication system modeled in Section \ref{SectionII}. Define $\mbox{\boldmath  $u$}[d]=[\mbox{\boldmath  $x$}[d-\nu+1], \dots, \mbox{\boldmath  $x$}[d]]$. It is easy to see $\mbox{\boldmath  $u$}(D)$ is a Markov sequence. The processed state $\mbox{\boldmath  $y$}[d]=\mbox{\boldmath  $y$}(\mbox{\boldmath  $u$}[d])$ is only a function of the corresponding Markov state. If two Markov states in successive time indices take the form
\begin{eqnarray}
&& \mbox{\boldmath  $u$}[d]=[\tilde{\mbox{\boldmath  $x$}}[d-\nu+1], \dots, \tilde{\mbox{\boldmath  $x$}}[d]] \nonumber \\
&& \mbox{\boldmath  $u$}[d+1]=[\tilde{\mbox{\boldmath  $x$}}[d-\nu+2], \dots, \tilde{\mbox{\boldmath  $x$}}[d+1]],
\end{eqnarray}
for some $\tilde{\mbox{\boldmath  $x$}}(D)$, then we have
\begin{equation}
P_t(\mbox{\boldmath  $u$}[d+1]|\mbox{\boldmath  $u$}[d])=\frac{1}{q^k}.
\end{equation}
Otherwise $P_t(\mbox{\boldmath  $u$}[d+1]|\mbox{\boldmath  $u$}[d])=0$. According to (\ref{PTR}), we have $p_{tr}=1$.

Since $\mbox{\boldmath  $u$}[d]=[\mbox{\boldmath  $x$}[d-\nu+1], \dots, \mbox{\boldmath  $x$}[d]]$ does not depend on source symbols at time indices $m \le d-\nu$, we know
\begin{equation}
P_t(\mbox{\boldmath  $u$}[d]|\mbox{\boldmath  $u$}[d-\nu])\ne 0, \qquad \forall \mbox{\boldmath  $u$}[d], \mbox{\boldmath  $u$}[d-\nu].
\end{equation}

The observation density is given by
\begin{equation}
f_o(\mbox{\boldmath  $r$}|\mbox{\boldmath  $y$})=\left(\frac{\mbox{\scriptsize SNR}}{2\pi}\right)^{\frac{n}{2}}\exp\left(-\frac{\mbox{\scriptsize SNR}}{2}\|\mbox{\boldmath  $r$}-g_q(\mbox{\boldmath  $y$} )\|^2 \right).
\end{equation}

Next, we show Assumption \ref{Assumption2} is satisfied. Let $\mbox{\boldmath  $u$}(D)$ and $\tilde{\mbox{\boldmath  $u$}}(D)$ be two Markov state sequences. Let $\mbox{\boldmath  $x$}(D)$ and $\mbox{\boldmath  $y$}(D)$ be the source message and the codeword corresponding to $\mbox{\boldmath  $u$}(D)$. Let $\tilde{\mbox{\boldmath  $x$}}(D)$ and $\tilde{\mbox{\boldmath  $y$}}(D)$ be the source message and the codeword corresponding to $\tilde{\mbox{\boldmath  $u$}}(D)$. For a time index $d$, if $\mbox{\boldmath  $u$}[d]\ne \tilde{\mbox{\boldmath  $u$}}[d]$, we can find a time index $m\in (d-\nu, d]$ such that $\mbox{\boldmath  $x$}[m]\ne \tilde{\mbox{\boldmath  $x$}}[m]$. Consequently, according to \cite[Corollary 2]{ref Forney73c}, we can find a time index $\tilde{m}\in [m, m+\nu)$, such that $\mbox{\boldmath  $y$}[\tilde{m}]\ne \tilde{\mbox{\boldmath  $y$}}[\tilde{m}]$. Therefore, Assumption \ref{Assumption2} holds because $\tilde{m}\in (d-\nu, d+\nu)$.

Let $d_{\min}^2$ and $d_{\max}^2$ be defined in Theorem \ref{Theorem1}. Let $\mbox{\boldmath  $y$}_1 \ne \mbox{\boldmath  $y$}_2$ be two arbitrary codeword symbols. We have the following triangle inequalities,
\begin{eqnarray}
&& \|\mbox{\boldmath  $r$}-g_q(\mbox{\boldmath  $y$}_2 )\| \ge \|g_q(\mbox{\boldmath  $y$}_2)-g_q(\mbox{\boldmath  $y$}_1 )\|- \|\mbox{\boldmath  $r$}-g_q(\mbox{\boldmath  $y$}_1 )\| \nonumber \\
&& \|\mbox{\boldmath  $r$}-g_q(\mbox{\boldmath  $y$}_2 )\| \le \|g_q(\mbox{\boldmath  $y$}_2)-g_q(\mbox{\boldmath  $y$}_1 )\|+ \|\mbox{\boldmath  $r$}-g_q(\mbox{\boldmath  $y$}_1 )\|.
\label{TriangleInequality}
\end{eqnarray}
The first inequality in (\ref{TriangleInequality}) implies
\begin{eqnarray}
\min_{\mbox{\scriptsize \boldmath  $y$}_2, \mbox{\scriptsize \boldmath  $y$}_2\ne \mbox{\scriptsize \boldmath  $y$}_1}\left[-\log(f_o(\mbox{\boldmath  $r$}|\mbox{\boldmath  $y$}_2))\right]+\log(f_o(\mbox{\boldmath  $r$}|\mbox{\boldmath  $y$}_1))  && =  \min_{\mbox{\scriptsize \boldmath  $y$}_2, \mbox{\scriptsize \boldmath  $y$}_2\ne \mbox{\scriptsize \boldmath  $y$}_1}\left[\frac{\mbox{\scriptsize SNR}}{2}(\|\mbox{\boldmath  $r$}-g_q(\mbox{\boldmath  $y$}_2 )\|^2-\|\mbox{\boldmath  $r$}-g_q(\mbox{\boldmath  $y$}_1 )\|^2) \right] \nonumber \\
&& \ge  \frac{\mbox{\scriptsize SNR}}{2}d_{\min}(d_{\min}-2 \|\mbox{\boldmath  $r$}-g_q(\mbox{\boldmath  $y$}_1 )\|).
\end{eqnarray}
The second inequality in (\ref{TriangleInequality}) implies
\begin{eqnarray}
\max_{\mbox{\scriptsize \boldmath  $y$}_2\ne \mbox{\scriptsize \boldmath  $y$}_3}\left[-\log(f_o(\mbox{\boldmath  $r$}|\mbox{\boldmath  $y$}_2))+\log(f_o(\mbox{\boldmath  $r$}|\mbox{\boldmath  $y$}_3))\right]  && = \max_{\mbox{\scriptsize \boldmath  $y$}_2\ne \mbox{\scriptsize \boldmath  $y$}_3}\left[\frac{\mbox{\scriptsize SNR}}{2}(\|\mbox{\boldmath  $r$}-g_q(\mbox{\boldmath  $y$}_2 )\|^2-\|\mbox{\boldmath  $r$}-g_q(\mbox{\boldmath  $y$}_3 )\|^2) \right] \nonumber \\
&& \le \max_{\mbox{\scriptsize \boldmath  $y$}_2}\left[\frac{\mbox{\scriptsize SNR}}{2}\|\mbox{\boldmath  $r$}-g_q(\mbox{\boldmath  $y$}_2 )\|^2\right] \nonumber \\
&& \le \max_{\mbox{\scriptsize \boldmath  $y$}_2}\left[\mbox{SNR}(\|\mbox{\boldmath  $r$}-g_q(\mbox{\boldmath  $y$}_1 )\|^2+\|g_q(\mbox{\boldmath  $y$}_2)-g_q(\mbox{\boldmath  $y$}_1 )\|^2)\right] \nonumber \\
&& \le \mbox{SNR}(\|\mbox{\boldmath  $r$}-g_q(\mbox{\boldmath  $y$}_1 )\|^2+d_{\max}^2).
\end{eqnarray}
Therefore, Assumption \ref{Assumption3} is satisfied by defining
\begin{eqnarray}
&& L_l(\mbox{\boldmath  $r$}, \mbox{\boldmath  $y$}_1)=\frac{\mbox{\scriptsize SNR}}{2}d_{\min}(d_{\min}-2 \|\mbox{\boldmath  $r$}-g_q(\mbox{\boldmath  $y$}_1 )\|) \nonumber \\
&& L_u(\mbox{\boldmath  $r$}, \mbox{\boldmath  $y$}_1)=\mbox{SNR}(\|\mbox{\boldmath  $r$}-g_q(\mbox{\boldmath  $y$}_1 )\|^2+d_{\max}^2 ).
\label{LowerandUpperBounds}
\end{eqnarray}
Note that evaluating $L_l(\mbox{\boldmath  $r$}, \mbox{\boldmath  $y$}_1)$ and $L_u(\mbox{\boldmath  $r$}, \mbox{\boldmath  $y$}_1)$ does not involve visiting any processed state other than $\mbox{\boldmath  $y$}_1$.

If $\mbox{\boldmath  $y$}[d]$ and $\mbox{\boldmath  $r$}[d]$ are the actual codeword symbol and the channel output at time index $d$, $\|\mbox{\boldmath  $r$}[d]-g_q(\mbox{\boldmath  $y$}[d])\|=\|\mbox{\boldmath  $n$}[d]\|$ is a $\chi^2$ random variable with mean $\frac{n}{\mbox{\scriptsize SNR}}$ and variance $\frac{2n}{{\mbox{\scriptsize SNR}}^2}$. From (\ref{LowerandUpperBounds}), it is easily seen that Assumption \ref{Assumption4} is satisfied with $d_{\min}^2>0$ and $d_{\max}^2<\infty$.

\subsection{Proof of Theorem \ref{Theorem3}}
\label{ProofTheorem3}

\begin{proof}
Let $\tilde{\mbox{\boldmath  $u$}}(D)$ be an arbitrary Markov state sequence with corresponding processed state sequence being $\tilde{\mbox{\boldmath  $y$}}(D)$. Assume
\begin{equation}
\tilde{\mbox{\boldmath  $u$}}[m+\nu-1] \ne \mbox{\boldmath  $u$}[m+\nu-1]
\label{ProofTheorem3.1}
\end{equation}
Theorem \ref{Theorem3} holds if we can prove that any $\tilde{\mbox{\boldmath  $u$}}(D)$ satisfying (\ref{ProofTheorem3.1}) cannot be the ML state sequence.

Let $k$ denote a positive integer. Define two integers $K_l$ and $K_r$ as follows.
\begin{eqnarray}
&& K_l= \mathop{\mbox{argmin}}_{k> 0} \{\tilde{\mbox{\boldmath  $u$}}[m+\nu-1-k\nu] = \mbox{\boldmath  $u$}[m+\nu-1-k\nu]\} \nonumber \\
&& K_r=\mathop{\mbox{argmin}}_{k> 0} \{ \tilde{\mbox{\boldmath  $u$}}[m+\nu-1+k\nu] = \mbox{\boldmath  $u$}[m+\nu-1+k\nu] \}.
\label{ProofTheorem3.2}
\end{eqnarray}
We consider respectively the following four cases based on the values of $K_l$ and $K_r$. In all the four cases, we show $\tilde{\mbox{\boldmath  $u$}}(D)$ cannot be the ML sequence.

{\bf Case 1: } $K_l\le 2M+1$, $K_r \le 2M-1$.

Since $\tilde{\mbox{\boldmath  $u$}}[m+\nu-1+k\nu] \ne \mbox{\boldmath  $u$}[m+\nu-1+k\nu]$ for all $-K_1<k<K_r$, according to Assumption \ref{Assumption2}, $\tilde{\mbox{\boldmath  $y$}}(D)$ and $\mbox{\boldmath  $y$}(D)$ differ at no less than $\left\lfloor\frac{K_l+K_r}{2}\right\rfloor$ time indices in the time interval $[m+\nu-K_l\nu, m+\nu+K_r\nu )$, where $\lfloor x \rfloor$ denotes the maximum integer no larger than $x$. According to (\ref{NLLu}) and (\ref{LBound}), for $d\in[m-2M\nu, m+2M\nu )$, if $\tilde{\mbox{\boldmath  $y$}}[d] \ne \mbox{\boldmath  $y$}[d]$, we have
\begin{equation}
-\log\frac{f_o(\mbox{\boldmath  $r$}[d]|\tilde{\mbox{\boldmath  $y$}}[d])}{f_o(\mbox{\boldmath  $r$}[d]|\mbox{\boldmath  $y$}[d])}\ge L_l(\mbox{\boldmath  $r$}[d], \mbox{\boldmath  $y$}[d]) > 3\nu(\rho-\log p_{tr}).
\end{equation}
Consequently, we get
\begin{eqnarray}
&& -\sum_{d=m+\nu-K_l\nu}^{m+\nu-1+K_r\nu}\log\frac{f_o(\mbox{\boldmath  $r$}[d]|\tilde{\mbox{\boldmath  $y$}}[d])P_t(\tilde{\mbox{\boldmath  $u$}}[d]|\tilde{\mbox{\boldmath  $u$}}[d-1])}{f_o(\mbox{\boldmath  $r$}[d]|\mbox{\boldmath  $y$}[d])P_t(\mbox{\boldmath  $u$}[d]|\mbox{\boldmath  $u$}[d-1])}  \nonumber \\
&& \ge \left\lfloor\frac{K_l+K_r}{2}\right\rfloor 3\nu(\rho-\log p_{tr})+ (K_r+K_l)\nu\log p_{tr}  \ge \left\lfloor\frac{K_l+K_r}{2}\right\rfloor 3\nu\rho >0
\label{ProofTheorem3.3}
\end{eqnarray}

According to the PCC presented in Appendix \ref{PCC}, (\ref{ProofTheorem3.3}) implies that $\mbox{\boldmath  $u$}(D)$ ``covers"\footnote{See definition in Appendix \ref{PCC}.} $\tilde{\mbox{\boldmath  $u$}}(D)$. Hence $\tilde{\mbox{\boldmath  $u$}}(D)$ cannot be the ML sequence.

{\bf Case 2: } $K_l\le 2M+1$, $K_r > 2M-1$.

In this case, we will construct a Markov sequence $\mbox{\boldmath  $u$}_c(D)$ and show that $\mbox{\boldmath  $u$}_c(D)$ covers $\tilde{\mbox{\boldmath  $u$}}(D)$.

$\mbox{\boldmath  $u$}_c(D)$ is constructed as follows.
\begin{eqnarray}
&& \mbox{\boldmath  $u$}_c[d]=\mbox{\boldmath  $u$}[d], \quad \mbox{for } d<m+2M\nu \nonumber \\
&& \mbox{\boldmath  $u$}_c[d]=\tilde{\mbox{\boldmath  $u$}}[d], \quad \mbox{for } d\ge m+(2M+1)\nu.
\label{ProofTheorem3.4}
\end{eqnarray}
According to (\ref{Homogeneous}), we can always construct $\mbox{\boldmath  $u$}_c[d]$ for $d\in [m+2M\nu, m+(2M+1)\nu)$ so that (\ref{ProofTheorem3.4}) is satisfied. Let $\mbox{\boldmath  $y$}_c(D)$ be the processed state sequence corresponding to $\mbox{\boldmath  $u$}_c(D)$.

From (\ref{NLLu}) and the first inequality in (\ref{LMBound}), we get
\begin{eqnarray}
&&  -\sum_{d=m+2M\nu}^{m+(2M+1)\nu}\log\frac{f_o(\mbox{\boldmath  $r$}[d]|\tilde{\mbox{\boldmath  $y$}}[d])P_t(\tilde{\mbox{\boldmath  $u$}}[d]|\tilde{\mbox{\boldmath  $u$}}[d-1])}{f_o(\mbox{\boldmath  $r$}[d]|\mbox{\boldmath  $y$}_c[d])P_t(\mbox{\boldmath  $u$}_c[d]|\mbox{\boldmath  $u$}_c[d-1])} \nonumber \\
&& \ge -\sum_{d=m+2M\nu}^{m+(2M+1)\nu-1}L_u(\mbox{\boldmath  $r$}[d], \mbox{\boldmath  $y$}[d])+(\nu+1)\log p_{tr}  \ge -3M\nu \rho
\label{ProofTheorem3.5}
\end{eqnarray}

Since $\tilde{\mbox{\boldmath  $u$}}[m+\nu-1+k\nu] \ne \mbox{\boldmath  $u$}_c[m+\nu-1+k\nu]$ for all $-K_l<k\le 2M-1$, according to Assumption \ref{Assumption2}, $\tilde{\mbox{\boldmath  $y$}}(D)$ and $\mbox{\boldmath  $y$}(D)$ differ at no less than $\left\lfloor\frac{K_l+2M-1}{2}\right\rfloor$ time indices in the time interval $[m+\nu-K_l\nu, m+2M\nu)$. According to (\ref{NLLu}) and (\ref{LBound}), we have
\begin{eqnarray}
&& -\sum_{d=m+\nu-K_l\nu}^{m+2M\nu-1}\log\frac{f_o(\mbox{\boldmath  $r$}[d]|\tilde{\mbox{\boldmath  $y$}}[d])P_t(\tilde{\mbox{\boldmath  $u$}}[d]|\tilde{\mbox{\boldmath  $u$}}[d-1])}{f_o(\mbox{\boldmath  $r$}[d]|\mbox{\boldmath  $y$}_c[d])P_t(\mbox{\boldmath  $u$}_c[d]|\mbox{\boldmath  $u$}_c[d-1])} \nonumber \\
&& >  \left\lfloor\frac{K_l+2M-1}{2}\right\rfloor 3\nu(\rho-\log p_{tr})  + (K_l+2M-1)\nu\log p_{tr} \nonumber \\
&& \ge  3M\nu(\rho-\log p_{tr}) + 2M\nu\log p_{tr} \ge  3M\nu\rho
\label{ProofTheorem3.6}
\end{eqnarray}

Combining (\ref{ProofTheorem3.5}) and (\ref{ProofTheorem3.6}), we obtain
\begin{equation}
-\sum_{d=m+\nu-K_l\nu}^{m+(2M+1)\nu}\log\frac{f_o(\mbox{\boldmath  $r$}[d]|\tilde{\mbox{\boldmath  $y$}}[d])P_t(\tilde{\mbox{\boldmath  $u$}}[d]|\tilde{\mbox{\boldmath  $u$}}[d-1])}{f_o(\mbox{\boldmath  $r$}[d]|\mbox{\boldmath  $y$}_c[d])P_t(\mbox{\boldmath  $u$}_c[d]|\mbox{\boldmath  $u$}_c[d-1])} >0
\label{ProofTheorem3.7}
\end{equation}
(\ref{ProofTheorem3.7}) implies that $\mbox{\boldmath  $u$}_c(D)$ covers $\tilde{\mbox{\boldmath  $u$}}(D)$. Hence according to the PCC, $\tilde{\mbox{\boldmath  $u$}}(D)$ cannot be the ML sequence.

{\bf Case 3: } $K_l> 2M+1$, $K_r \le 2M-1$.

Similar to Case 2, we will construct a Markov sequence $\mbox{\boldmath  $u$}_c(D)$ and show that $\mbox{\boldmath  $u$}_c(D)$ covers $\tilde{\mbox{\boldmath  $u$}}(D)$.

$\mbox{\boldmath  $u$}_c(D)$ is constructed as follows.
\begin{eqnarray}
&& \mbox{\boldmath  $u$}_c[d]=\mbox{\boldmath  $u$}[d], \quad \mbox{for } d\ge m-2M\nu \nonumber \\
&& \mbox{\boldmath  $u$}_c[d]=\tilde{\mbox{\boldmath  $u$}}[d], \quad \mbox{for } d < m-(2M+1)\nu.
\label{ProofTheorem3.8}
\end{eqnarray}
According to (\ref{Homogeneous}), we can always construct $\mbox{\boldmath  $u$}_c[d]$ for $d\in [m-(2M+1)\nu, m-2M\nu)$ so that (\ref{ProofTheorem3.8}) is satisfied. Let $\mbox{\boldmath  $y$}_c(D)$ be the processed state sequence corresponding to $\mbox{\boldmath  $u$}_c(D)$.

From (\ref{NLLu}) and the second inequality in (\ref{LMBound}), we get
\begin{eqnarray}
&&  -\sum_{d=m-(2M+1)\nu}^{m-2M\nu-1}\log\frac{ f_o(\mbox{\boldmath  $r$}[d]|\tilde{\mbox{\boldmath  $y$}}[d])P_t(\tilde{\mbox{\boldmath  $u$}}[d]|\tilde{\mbox{\boldmath  $u$}}[d-1])}{f_o(\mbox{\boldmath  $r$}[d]|\mbox{\boldmath  $y$}_c[d])P_t(\mbox{\boldmath  $u$}_c[d]|\mbox{\boldmath  $u$}_c[d-1])} \nonumber \\
&& \ge -\sum_{d=m-(2M+1)\nu}^{m-2M\nu-1}L_u(\mbox{\boldmath  $r$}[d], \mbox{\boldmath  $y$}[d])+\nu\log p_{tr}  \ge -3M\nu \rho.
\label{ProofTheorem3.9}
\end{eqnarray}

Since $\tilde{\mbox{\boldmath  $u$}}[m+\nu-1+k\nu] \ne \mbox{\boldmath  $u$}_c[m+\nu-1+k\nu]$ for all $-2M-1 \le k <K_r $, according to Assumption \ref{Assumption2}, $\tilde{\mbox{\boldmath  $y$}}(D)$ and $\mbox{\boldmath  $y$}(D)$ differ at no less than $\left\lfloor\frac{2M+1+K_r}{2}\right\rfloor$ time indices in the time interval $[m-2M\nu, m+\nu+K_r\nu)$. According to (\ref{NLLu}) and (\ref{LBound}), we have
\begin{eqnarray}
&& -\sum_{d=m-2M\nu}^{m+\nu+K_r\nu-1}\log\frac{ f_o(\mbox{\boldmath  $r$}[d]|\tilde{\mbox{\boldmath  $y$}}[d])P_t(\tilde{\mbox{\boldmath  $u$}}[d]|\tilde{\mbox{\boldmath  $u$}}[d-1])}{f_o(\mbox{\boldmath  $r$}[d]|\mbox{\boldmath  $y$}_c[d])P_t(\mbox{\boldmath  $u$}_c[d]|\mbox{\boldmath  $u$}_c[d-1])} \nonumber \\
&& > \left\lfloor\frac{2M+1+K_r}{2}\right\rfloor 3\nu(\rho-\log p_{tr})   + (2M+1+K_r)\nu\log p_{tr} \ge 3(M+1)\nu\rho.
\label{ProofTheorem3.10}
\end{eqnarray}

Combining (\ref{ProofTheorem3.9}) and (\ref{ProofTheorem3.10}), we obtain
\begin{equation}
\sum_{d=m-(2M+1)\nu}^{m+\nu+K_r\nu-1}\log\frac{ f_o(\mbox{\boldmath  $r$}[d]|\tilde{\mbox{\boldmath  $y$}}[d])P_t(\tilde{\mbox{\boldmath  $u$}}[d]|\tilde{\mbox{\boldmath  $u$}}[d-1])}{f_o(\mbox{\boldmath  $r$}[d]|\mbox{\boldmath  $y$}_c[d])P_t(\mbox{\boldmath  $u$}_c[d]|\mbox{\boldmath  $u$}_c[d-1])} < 0
\label{ProofTheorem3.11}
\end{equation}
(\ref{ProofTheorem3.11}) implies that $\mbox{\boldmath  $u$}_c(D)$ covers $\tilde{\mbox{\boldmath  $u$}}(D)$. Hence according to the PCC, $\tilde{\mbox{\boldmath  $u$}}(D)$ cannot be the ML sequence.

{\bf Case 4: } $K_l> 2M+1$, $K_r > 2M-1$.

We construct a Markov state sequence $\mbox{\boldmath  $u$}_c(D)$ as follows.
\begin{eqnarray}
&& \mbox{\boldmath  $u$}_c[d]=\mbox{\boldmath  $u$}[d], \quad \mbox{for } m-2M\nu \le d<m+2M\nu \nonumber \\
&& \mbox{\boldmath  $u$}_c[d]=\tilde{\mbox{\boldmath  $u$}}[d], \quad \mbox{for } d\ge m+(2M+1)\nu \nonumber \\
&& \mbox{\boldmath  $u$}_c[d]=\tilde{\mbox{\boldmath  $u$}}[d], \quad \mbox{for } d < m-(2M+1)\nu.
\label{ProofTheorem3.12}
\end{eqnarray}
Let the processed state sequence corresponding to $\mbox{\boldmath  $u$}_c(D)$ be $\mbox{\boldmath  $y$}_c(D)$.

Since $\tilde{\mbox{\boldmath  $u$}}[m+\nu-1+k\nu] \ne \mbox{\boldmath  $u$}_c[m+\nu-1+k\nu]$ for all $-2M-1 \le k \le 2M-1 $, according to Assumption \ref{Assumption2}, $\tilde{\mbox{\boldmath  $y$}}(D)$ and $\mbox{\boldmath  $y$}(D)$ differ at no less than $\left\lfloor\frac{4M+1}{2}\right\rfloor$ time indices in the time interval $[m-2M\nu, m+2M\nu)$. According to (\ref{NLLu}) and (\ref{LBound}), we have
\begin{equation}
-\sum_{d=m-2M\nu}^{m+2M\nu-1}\log\frac{f_o(\mbox{\boldmath  $r$}[d]|\tilde{\mbox{\boldmath  $y$}}[d])P_t(\tilde{\mbox{\boldmath  $u$}}[d]|\tilde{\mbox{\boldmath  $u$}}[d-1])}{f_o(\mbox{\boldmath  $r$}[d]|\mbox{\boldmath  $y$}_c[d])P_t(\mbox{\boldmath  $u$}_c[d]|\mbox{\boldmath  $u$}_c[d-1])} > \left\lfloor\frac{4M+1}{2}\right\rfloor 3\nu(\rho-\log p_{tr}) + 4M\nu\log p_{tr} \ge 6M\nu\rho.
\label{ProofTheorem3.13}
\end{equation}

Meanwhile, it is easily seen that (\ref{ProofTheorem3.5}) and (\ref{ProofTheorem3.9}) hold. Combine (\ref{ProofTheorem3.5}), (\ref{ProofTheorem3.9}) and (\ref{ProofTheorem3.13}), we obtain
\begin{eqnarray}
&& -\sum_{d=m-(2M+1)\nu}^{m+(2M+1)\nu}\log\frac{f_o(\mbox{\boldmath  $r$}[d]|\tilde{\mbox{\boldmath  $y$}}[d]))P_t(\tilde{\mbox{\boldmath  $u$}}[d]|\tilde{\mbox{\boldmath  $u$}}[d-1])}{f_o(\mbox{\boldmath  $r$}[d]|\mbox{\boldmath  $y$}_c[d])P_t(\mbox{\boldmath  $u$}_c[d]|\mbox{\boldmath  $u$}_c[d-1])}  > -3M\nu \rho -3M\nu \rho +6M\nu\rho = 0.
\label{ProofTheorem3.14}
\end{eqnarray}
(\ref{ProofTheorem3.14}) implies that $\mbox{\boldmath  $u$}_c(D)$ covers $\tilde{\mbox{\boldmath  $u$}}(D)$. Hence according to the PCC, $\tilde{\mbox{\boldmath  $u$}}(D)$ cannot be the ML sequence.

Overall, we showed that $\tilde{\mbox{\boldmath  $u$}}(D)$ cannot be the ML sequence irrespective of the values of $K_l$ and $K_r$. Therefore, $\tilde{\mbox{\boldmath  $u$}}[m+\nu-1] = \mbox{\boldmath  $u$}[m+\nu-1]$ must be true.
\end{proof}

\end{document}